\documentclass{article}

\usepackage{arxiv}

\usepackage[utf8]{inputenc}
\usepackage[T1]{fontenc}
\usepackage{amsmath}
\usepackage{hyperref}
\usepackage{cleveref}
\usepackage{url}
\usepackage{booktabs}
\usepackage{amsfonts}
\usepackage{nicefrac}
\usepackage{microtype}
\usepackage{lipsum}
\usepackage{graphicx}
\graphicspath{ {./images/} }

\title{On the application of Visibility Graphs in the Spectral Domain for Speaker Recognition}

\author{
  Hernan Bocaccio \\
  Departamento de Física \\ 
  Universidad de Buenos Aires \\
  INFINA - CONICET \\
  Buenos Aires, Argentina \\
  \And
  Sergio Iglesias-Pérez \\
  Departamento de Matemática Aplicada \\ 
  Universidad Rey Juan Carlos \\
  Madrid, España \\
  \And
  Miguel Romance \\
  Departamento de Matemática Aplicada \\ 
  Universidad Rey Juan Carlos \\
  Madrid, España \\
  \And
  Regino Criado \\
  Departamento de Matemática Aplicada \\ 
  Universidad Rey Juan Carlos \\
  Madrid, España \\
  \And
  Gabriel B. Mindlin \\
  Departamento de Física \\ 
  Universidad de Buenos Aires \\
  INFINA - CONICET \\
  Buenos Aires, Argentina \\
  }

\begin{document}
\maketitle
\begin{abstract}
In this study, we explore the potential of visibility graphs in the spectral domain for speaker recognition. Adult participants were instructed to record vocalizations of the five Spanish vowels. For each vocalization, we computed the frequency spectrum considering the source-filter model of speech production, where formants are shaped by the vocal tract acting as a passive filter with resonant frequencies. Spectral profiles exhibited consistent intra-speaker characteristics, reflecting individual vocal tract anatomies, while showing variation between speakers. We then constructed visibility graphs from these spectral profiles and extracted various graph-theoretic metrics to capture their topological features. These metrics were assembled into feature vectors representing the five vowels for each speaker. Using an ensemble of decision trees trained on these features, we achieved high accuracy in speaker identification. Our analysis identified key topological features that were critical in distinguishing between speakers. This study demonstrates the effectiveness of visibility graphs for spectral analysis and their potential in speaker recognition. We also discuss the robustness of this approach, offering insights into its applicability for real-world speaker recognition systems. This research contributes to expanding the feature extraction toolbox for speaker recognition by leveraging the topological properties of speech signals in the spectral domain.
\end{abstract}


\section{Introduction}
Visibility graphs were developed not many years ago, offering the representation of temporal signals as complex networks\cite{lacasa_time_2008}. Since then, they have been established as a powerful tool in network analysis and have demonstrated versatility in widespread applications across various disciplines and topics in science. The conversion of time series to visibility graphs has proven to be a powerful approach for capturing complex temporal properties in signal analysis, as it effectively identifies intricate relationships and patterns within temporal sequences by abstracting temporal dynamics into graphical structures \cite{lacasa_visibility_2009, luque_horizontal_2009}. This development has opened up numerous possibilities for approaching signal processing from a topological perspective, in contrast to the more commonly used methods from algebra and differential calculus. Visibility graphs have been employed to unravel intricate temporal phenomena, with applications in the study of chaotic behavior in cryptocurrency prices \cite{partida_chaotic_2022}, network intrusion detection systems for cybersecurity by analyzing the dynamics of interactions between computers \cite{iglesias-perez_increasing_2022}, the relationship between real estate prices and citizen commuting dynamics \cite{iglesias-perez_temporal_2023, iglesias-perez_combining_2023}, biological state differentiation based on biomarker dynamics across various biological datasets \cite{zheng_visibility_2021}, synchronization in coupled dynamical systems \cite{ahmadlou_visibility_2012}, the extraction of hierarchical rhythmic self-similarities in local standard deviations series from audio signals for music genre classification \cite{melo_graph-based_2020}, and temporal dynamics in macroscale brain activity related to disorders \cite{sulaimany_visibility_2023}. Furthermore, visibility graphs have been shown to be extendable to the analysis of spatial patterns in scalar fields \cite{lacasa_visibility_2017}, which can be readily adapted for image processing \cite{iacovacci_visibility_2020, pei_texture_2021} and image classification \cite{wen_visibility_2022}. Hence, visibility graphs have demonstrated versatile potential applications for studying complex patterns in structured data in general.

Particularly intriguing is the potential application of visibility graphs to spectral series, offering a novel approach to analyzing frequency-domain data \cite{yela_spectral_2019}. Leveraging visibility graphs, due to their ability to detect periodicity through peak identification even in noisy signals \cite{nunez_detecting_2012}, allows for the capture of harmonic content in spectral domain series while filtering out components associated with noise. These noise components, often spread across a wider frequency range at a given time point, are effectively rejected. This approach preserves the relevant spectral information associated with harmonics, such as those found in music and speech, while minimizing the impact of noise from other sources \cite{yela_spectral_2019}. As a result, visibility graphs present clear advantages over traditional methods for identifying harmonic peaks and calculating the intervals between them.

Thus, this approach appears to be a valuable tool for studying human vocalizations within the framework of the source-filter theory of vocal production \cite{fant_acoustic_1971, lieberman_speech_1988, titze_principles_1994}. According to this theory, the modulation of airflow by the vocal folds serves as the source of voiced sounds, which contain spectrally rich content. This airflow is subsequently shaped by the vocal tract, acting as a passive filter, where resonant frequencies, known as formants, are highlighted. As a result, the spectral richness of the sound depends on the anatomical characteristics of the speaker, functioning as unique voiceprints with distinct properties \cite{trevisan_topological_2005}, which can then be used for speaker identification.

In this study, we investigate speaker recognition using spectral graph analysis of vowel vocalizations in the Spanish language. Participants were instructed to produce recordings repeating vocalizations of each of the five Spanish vowels. The recordings were preprocessed and segmented into individual vowel vocalizations. From each segment, we calculated the formants of speech, by using Linear Programming Coefficients (LPC), as a representation of the frequency spectrum characterizing each vocalization. We observed that the spectral profiles exhibited similar characteristics within individual subjects, reflecting the anatomical properties of their vocal tracts, while differing between subjects. By applying visibility graph analysis to these spectral representations, we transform them into graphs capturing the topological properties of the vowel spectra. Leveraging graph theory metrics, we construct attribute vectors encapsulating the topological characteristics of each vowel. These metrics were combined into attribute vectors representing a set of five vocalizations, one for each vowel. Utilizing an ensemble of decision trees by a Random Forest model, trained on these attribute vectors, we achieve high accuracy in predicting the identity of speakers. Furthermore, employing feature importance analysis, we identify key topological features crucial for speaker discrimination. Present work results demonstrate the feasibility of speaker recognition based on topological metrics derived from spectral graphs of vowel vocalizations in the Spanish language. Our findings not only underscore the efficacy of visibility graphs as a powerful tool in the spectral domain but also demonstrate their utility in speaker recognition tasks. This application is particularly significant in various fields, including cybersecurity, where accurate speaker identification is crucial. This study not only contributes to advancing speaker recognition methodologies but also highlights the broader potential of graph-based approaches in analyzing spectral data across various domains.

\section{Methods}

\subsection{Data acquisition and preprocess}

We collected data from 7 male adult subjects, with ages in the range of (37.86 $\pm$ 5.15) years. Participants were instructed to use the microphone on their phones to record audio samples articulating the five Spanish vowels in vowel-separated audio recordings. In each recording session, they were directed to produce isolated vocalizations of each specific Spanish vowel, repeating it multiple times for at least half a minute. As a result, each audio file contains various samples of vocalizations of a single unique vowel, and the complete set of recordings constitutes a comprehensive collection of vocalizations representing the five vowels that comprise the Spanish language. All subjects are native Spanish speakers. The recordings were requested to be performed in a controlled environment to minimize background noise and ensure recording consistency.

The audio signals were first converted to mono in WAV format and resampled to a uniform sampling rate of 11025 Hz to ensure consistent data processing across all recordings. This sampling rate was chosen for its computational efficiency while preserving the relevant frequency content necessary for formant analysis. The recordings were then preprocessed to eliminate artifacts and residual background noise using spectral gating noise reduction techniques \cite{sainburg_finding_2020}. The preprocessed audio files were segmented using the $split\_on\_silence$ function from the Pydub Python package, which divides audio into smaller chunks by detecting pauses or silences based on a specified threshold for duration and volume. After careful review, artifactual segments were manually discarded, ensuring that each segment contained only a single vocalization of one of the five vowels. This allowed us to isolate each vowel sound for subsequent spectral function computation, resulting in a total of 890 stratified audio segments (see \Cref{table1}).

\begin{table}[!ht]
 \caption{\label{table1}
 Audio segments for each subject and vowel.}
  \centering
  \begin{tabular}{cccccc}
    \toprule
    \toprule
     & $\backslash a \backslash$ & $\backslash e \backslash$ & $\backslash i \backslash$ & $\backslash o \backslash$ & $\backslash u \backslash$ \\
    \midrule
    \midrule
    S01 & 35 & 37 & 33 & 34 & 34 \\
    S02 & 22 & 24 & 26 & 28 & 27 \\
    S03 & 24 & 25 & 25 & 25 & 26 \\
    S04 & 19 & 22 & 25 & 25 & 31 \\
    S05 & 22 & 22 & 24 & 24 & 24 \\
    S06 & 25 & 26 & 27 & 26 & 26 \\
    S07 & 21 & 19 & 19 & 19 & 19 \\
    \bottomrule
    \bottomrule
  \end{tabular}
  \label{tab:table}
\end{table}

\subsection{Spectral functions}
 
For the extraction of the spectral profile, we compute the speech formants by employing a linear predictive coding (LPC) approach, a well-established method in speech processing for modeling the vocal tract filter related spectral content of the speech signals \cite{atal_speech_1971}. The LPC coefficients were computed using the Librosa package of Python \cite{mcfee_librosa_2015}, with the LPC order set to 13. This selection follows the methodology of previous works \cite{trevisan_topological_2005}, where 13 poles were sufficient to capture the main features of the formant’s envelope, offering a balance between model complexity and the accuracy of the spectral representation of the vowels. Once the LPC coefficients were obtained, we computed the frequency response of the system using the $scipy.signal.freqz$ function \cite{virtanen_scipy_2020}, specifying the use of 512 discrete frequency bins to ensure a high-resolution spectral profile. Hence, the power spectral function was estimated according \Cref{eq1}, using $d_0=1$ and $d_k$ as the set of LPC coefficients.

\begin{equation}
H(f) = \frac{d_{0}}{1-\sum_{k=1}^{m} d_{k}e^{i k 2 \pi f \Delta}}
\label{eq1}
\end{equation}

Due to the sampling rate after resampling, the spectral profiles were computed over a frequency range of 0 to approximately 5512 Hz (half the sampling rate), which encompasses the frequency range of interest for the vowels under analysis. In \Cref{method}a we show an example of a log power spectral function, i.e. $log(|H(f)|^2)$, where $H$ is the frequency response computed in \Cref{eq1}, extracted from an isolated audio. We also show the associated spectrogram, reflecting the spotlight of resonant frequencies characteristics through the spectral profile representation. In addition to the analysis conducted using an LPC order of 13, we performed a sensitivity analysis by calculating spectral profiles for a range of LPC orders, from 10 to 20. This exploration aimed to evaluate the robustness of the method and the sensitivity of the corresponding analysis to the chosen LPC order.

\subsection{Selection of representative spectra}

To proceed with the analysis, we identified the most representative spectral profiles for each speaker and vowel. This was accomplished through a community detection approach based on spectral similarity. Initially, we calculated the pairwise correlations between the log power spectral functions. These correlations were then binarized using a threshold of 0.9, resulting in an undirected adjacency matrix. We utilized the BCT toolbox \cite{rubinov_complex_2010} to perform community detection on this matrix, resulting in a separation of the spectral functions into distinct communities. From this, we selected the largest community, also known as the giant component, which we considered to represent the most characteristic spectral patterns for each subject and vowel. The spectral profiles associated with this largest component were about 83\% of the total and were used for all subsequent analyses.

While the primary results were obtained using a correlation threshold of 0.9, we further tested the robustness of our method by repeating the entire process with threshold values ranging from 0.5 to 0.95, in increments of 0.05. This approach allowed us to assess the stability of our results across different threshold choices and evaluate the sensitivity of the spectral functions labeled as equivalent. As a result, we demonstrated the method's potential for generalization, even in the presence of spectral variance caused by different sources of degradation.

\subsection{Visibility graph and graph-based features}

\begin{figure}[!ht]
    \centering
    \includegraphics[scale=0.55]{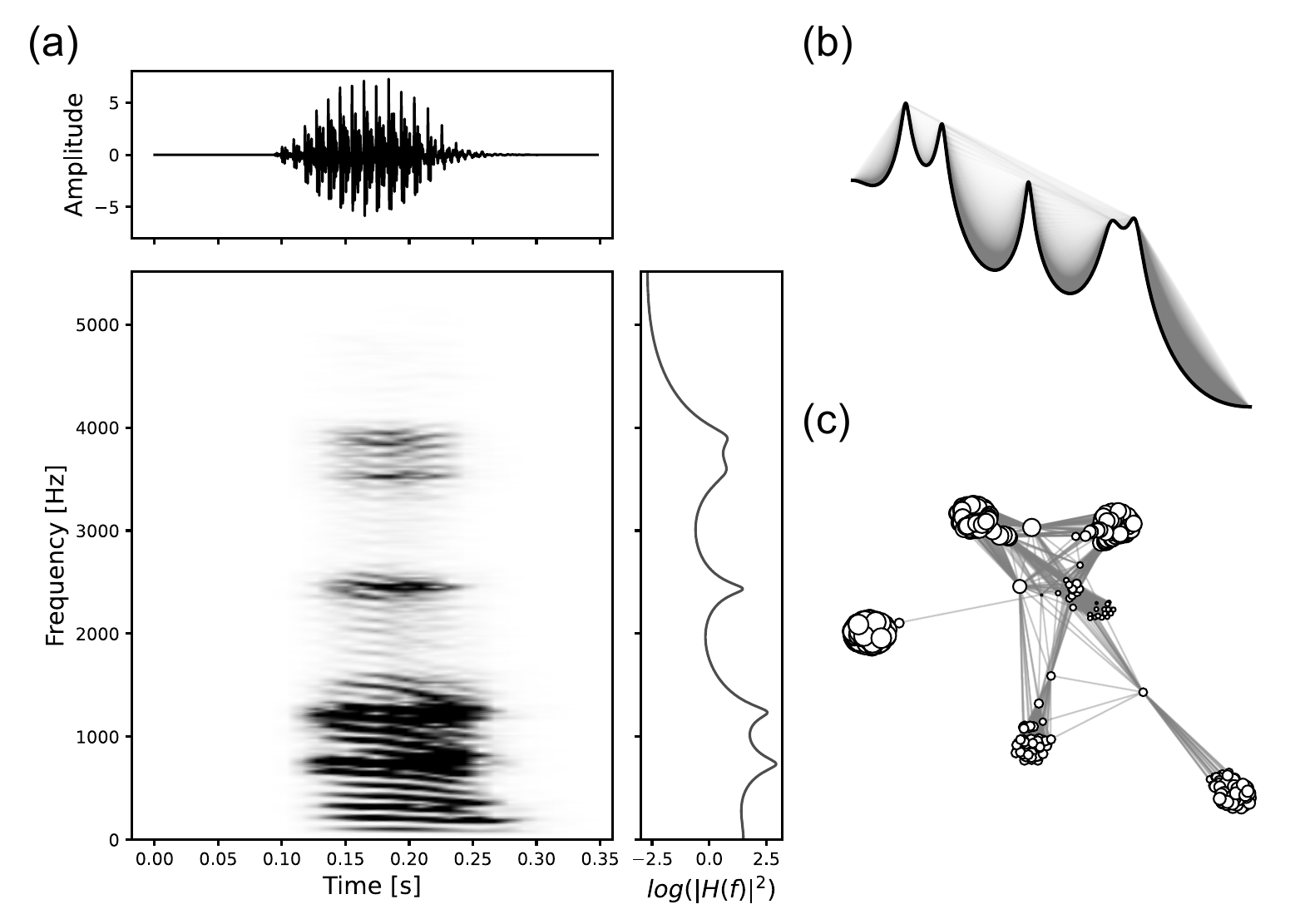}
    \caption{\label{method} Example of an audio segment data, with its corresponding amplitude across time, spectrogram, and the spectrum profile obtained from it shown in (a). From this spectrum profile we computed: (b) the connections between the elements of the temporal series according to \Cref{eq2}; (c) the natural visibility graph associated in a force-based layout.}
\end{figure}

We constructed the visibility graphs from spectral profiles descriptive below. For this purpose, we used the definition of natural visibility graphs \cite{lacasa_time_2008} according to \Cref{eq2}. This involved transforming the spectral representations of each vowel segment into visibility graphs, where nodes represent the spectral profile amplitude of discrete frequency components and edges denote visibility relationships between nodes according to this amplitude. We show the visibility links computed between discrete frequencies of the series in the spectral domain for the previous audio example (\Cref{method}b) and the resulting visibility graph with a forced-based layout (\Cref{method}c). The construction of visibility graphs allowed us to abstract the spectral data into graphical structures, facilitating the analysis of connectivity patterns and topological features embedded within the frequency spectra. We applied a divide and conquer strategy for faster algorithm procedures \cite{lan_fast_2015} which has shown the fastest offline computing algorithm for general time series \cite{yela_online_2020}.

\begin{equation}
y_{c} < y_{b} + (y_{a}-y_{b})\frac{t_{b}-t_{c}}{t_{b}-t_{a}}
\label{eq2}
\end{equation}

Once the visibility graphs were constructed, we applied graph theory metrics to quantify various topological properties of the graphs. The metrics included were link density, average path length, clustering coefficient, and modularity. By computing these metrics for each vowel segment, we generated attribute vectors encapsulating the topological characteristics of the spectral data. These attribute vectors served as input features for subsequent speaker recognition analysis.

\newpage

\subsection{Data split and features construction}

The acoustic data segments for each subject and vowel were randomly divided into training (40\%), validation (30\%), and test (30\%) sets. For this data split, we combined the metrics associated with the five vowels, randomly selecting segments of each vowel while ensuring that the combinations were made within each set to prevent data leakage. Thus, for each combination, we obtained a feature array containing the graph-based metrics from one example audio segment for each vowel, which could then be assigned to a corresponding subject label. We performed a total of 1000 combinations per subject for each data set, ensuring that this value was lower than the total number of possible combinations based on the available segments. These combined feature arrays were then used as inputs to the model throughout the different stages of the process.

\subsection{Speaker recognition model training, evaluation, and feature importance analysis}

To train and evaluate our speaker recognition model, we performed 10 runs with different data splitting and feature combination processes. For the model, we employed an ensemble of decision trees using the Random Forest algorithm \cite{breiman_random_2001}. The attribute vectors derived from the graph-theory metrics served as input features, while the target labels corresponded to the identities of the speakers. We trained the models using a supervised learning framework. We optimized the model hyperparameters through the maximization of the accuracy in the validation set, to reduce overfitting. We performed a grid search of hyperparameters across variations in the number of estimators ($n\_estimators$) ranging between 5 and 50 with steps of 5, and the maximum depth of the trees ($max\_depth$) ranging between 5 and 15 with steps of 1. We selected optimal parameters based on the validation scores obtained during the grid search. Then models were trained for this fine-tuned hyperparameters across the development set data (train and validation). To assess the performance of our speaker recognition model, we used standard performance metrics such as precision, recall, and F1-score. Additionally, we performed feature importance analysis using Random Forest models to identify the most discriminative topological properties for speaker recognition. We computed Shapley values using SHapley Additive exPlanation (SHAP) \cite{lundberg_unified_2017} with an efficient tree-based implementation \cite{lundberg_local_2020} on the test set. The interventional method was applied to break feature correlations, ensuring a more accurate estimation of each attribute’s contribution \cite{janzing_feature_2020}. This analysis provided insights into the spectral patterns captured by visibility graphs, reinforcing their utility for speaker identification.

\section{Results} 

\subsection{Spectral analysis of vocalizations per speaker}

\begin{figure}[!ht]
    \centering
    \includegraphics[scale=0.40]{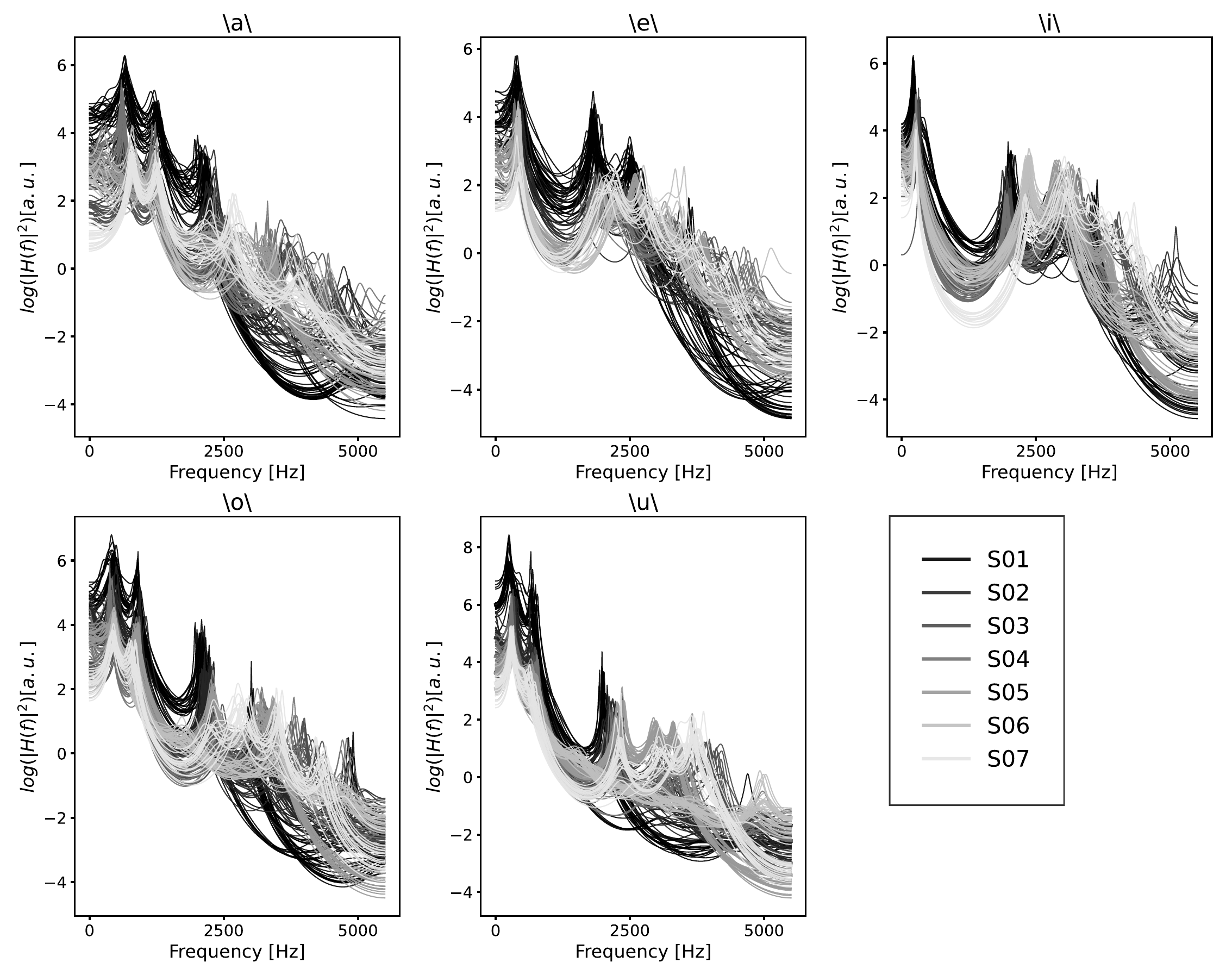}
    \caption{\label{spectrum} Spectrum functions magnitude of all subjects through the vocalization of the five different spanish vowels.}
\end{figure}

We first performed the spectral analysis of vocalizations per speaker. \Cref{spectrum} presents the representative spectral profiles of vocalizations for each of the five vowels ($\backslash a \backslash$, $\backslash e \backslash$, $\backslash i \backslash$, $\backslash o \backslash$, $\backslash u \backslash$) produced by the seven speakers (S01 to S07). The logarithmic power spectral functions ($log(|H(f)|^2)$), are plotted against frequency for each vowel across all speakers. Each line within a subplot corresponds to a single vocalization, color-coded according to the speaker, with darker lines representing lower-numbered speakers and progressively lighter lines for higher-numbered speakers.

For all vowels, the spectral profiles exhibit characteristic peaks corresponding to formant frequencies, which are influenced by the physiological properties of the speaker's vocal tract. Despite the shared vowel identity, noticeable inter-speaker variability is observed in the spectral contours. These differences are most pronounced in the higher frequency ranges (above 2000 Hz), where individual speaker characteristics, such as vocal tract length and articulation dynamics, contribute to the variability. In contrast, lower-frequency components (below 1000 Hz) display more uniformity across speakers for the same vowel, reflecting the common vowel articulation properties, particularly related to the fundamental frequency and first formant.

\newpage

\subsection{Graph-based metrics across speakers}

\begin{figure}[!ht]
    \centering
    \includegraphics[scale=0.5]{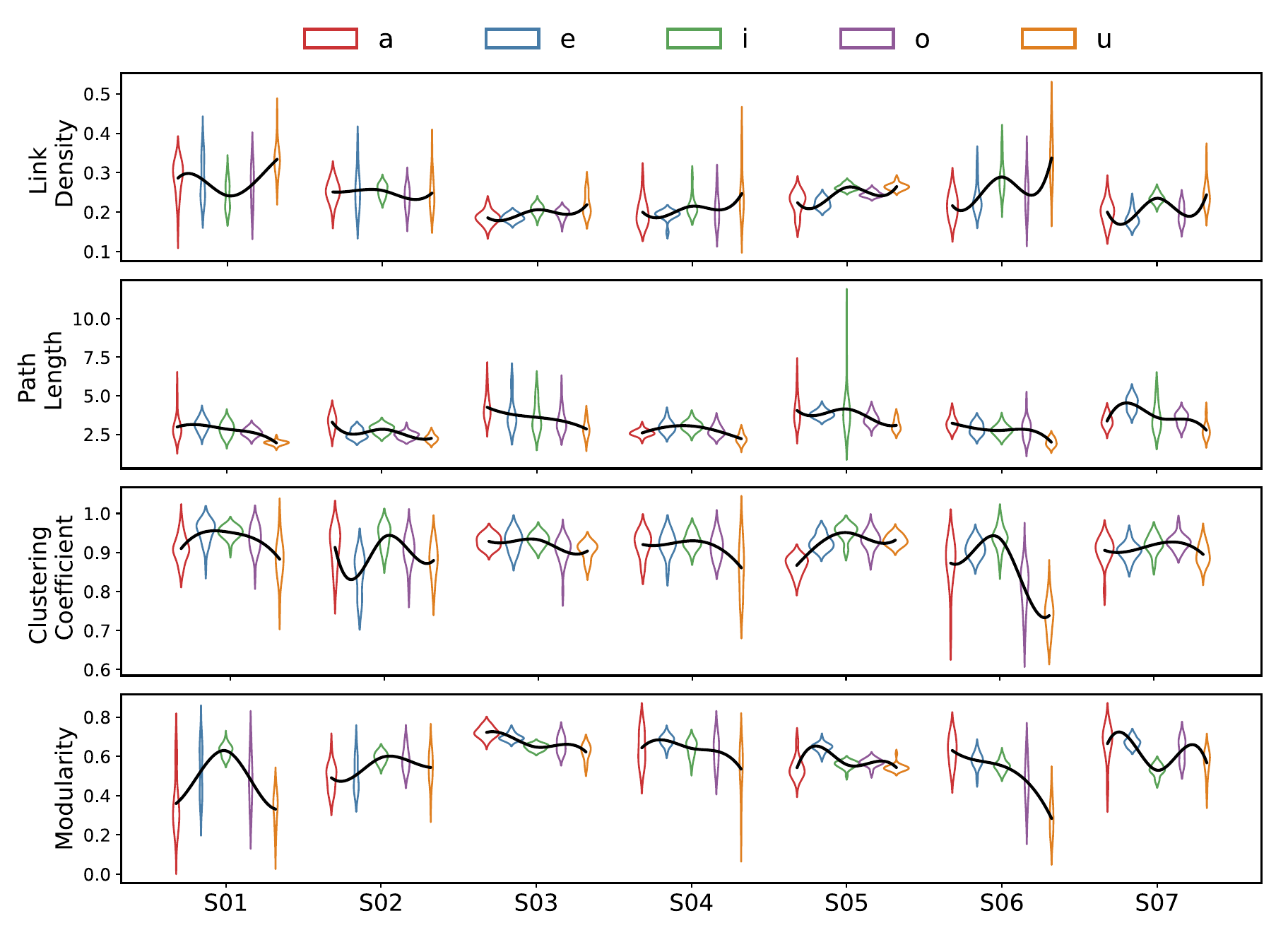}
    \caption{\label{metrics} Graph-based metrics distributions obtained from the visibility graphs of all audio segments, grouped by subjects and vowels.}
\end{figure}

The distributions obtained for the four graph-based metrics (link density, average path length, clustering coefficient, and modularity) calculated for the visibility graphs derived from the representative spectral profiles of the five vowels for each speaker are illustrated in \Cref{metrics}. The colored violin plots represent the distribution of each metric for the corresponding vowel, while the black lines connecting the means of these distributions provide a smoothed representation of the general trend across vowels. The trends captured by these metrics reveal vowel-specific patterns that vary across speakers. The combination of graph-theoretical features provides complementary information that could help distinguish speakers, as evidenced by the distinct patterns formed by the smoothed mean lines. These results suggest that the use of these metrics from all five vowels could give high discriminatory capabilities of the method for speaker recognition.

\subsection{Speaker recognition}

We trained the models using 10 independent random splits of the training and test datasets. For each model, hyperparameter fine-tuning resulted in optimal configurations involving relatively shallow trees (i.e., $max\_depth$ values between 5 and 10) combined with a moderate to high number of estimators (ranging from 20 to 50). These optimal settings were consistent across different speakers, indicating that similar model complexities are effective for distinguishing between individuals. This consistency suggests that the graph-theoretical features derived from visibility graphs capture speaker-specific characteristics that can be effectively utilized by models of relatively low complexity.

Speaker recognition models exhibited remarkable performances in identifying individual speakers. We evaluated the model's performance on the test set using precision, recall, and F1-score. The results demonstrated high levels of performances across all speaker classes, underscoring the robustness and effectiveness of our approach in distinguishing between different speakers based on spectral features. \Cref{performance}a provides a detailed evaluation of the model’s precision, recall, and F1-score across all speakers (S01 to S07) and compares the model's performance on the actual data versus a randomized control, obtained by permuting target labels. The left side of the panel shows the precision, recall, and F1-score for each speaker, represented by black circles for the actual data and hollow circles for the randomized data. For the actual data, precision, recall, and F1-scores remain consistently high across all speakers, with values close to or at 1.0 for most individuals. This indicates that the model was able to accurately classify vocalizations, rarely mislabeling one speaker’s data as another’s (high precision) and demonstrating a high rate of correctly identified true positives (high recall). The consistently high F1-scores, which provide a harmonic mean of precision and recall, underscore the balance of these two metrics across the dataset. The performance obtained on randomized data is significantly lower, with precision, recall, and F1-scores around 0.2. This substantial drop when using random data confirms the model’s reliance on meaningful patterns inherent in the topological features extracted from the visibility graphs. The right side of the panel provides a summary of macro-averaged precision, recall, and F1-scores for both the actual and randomized data. The macro-averaged values for the actual data consistently approach 1.0 ($\text{Precisión} = 0.951 \pm 0.028$; $\text{Recall} = 0.948 \pm 0.030$; $\text{F1-score} = 0.947 \pm 0.030$), while those for the random data remain low, reinforcing the model’s capacity to generalize effectively across the dataset and emphasizing the significance of the topological features in distinguishing speakers.

\subsection{Feature importance analysis}

\begin{figure}[!ht]
    \centering
    \includegraphics[scale=0.55]{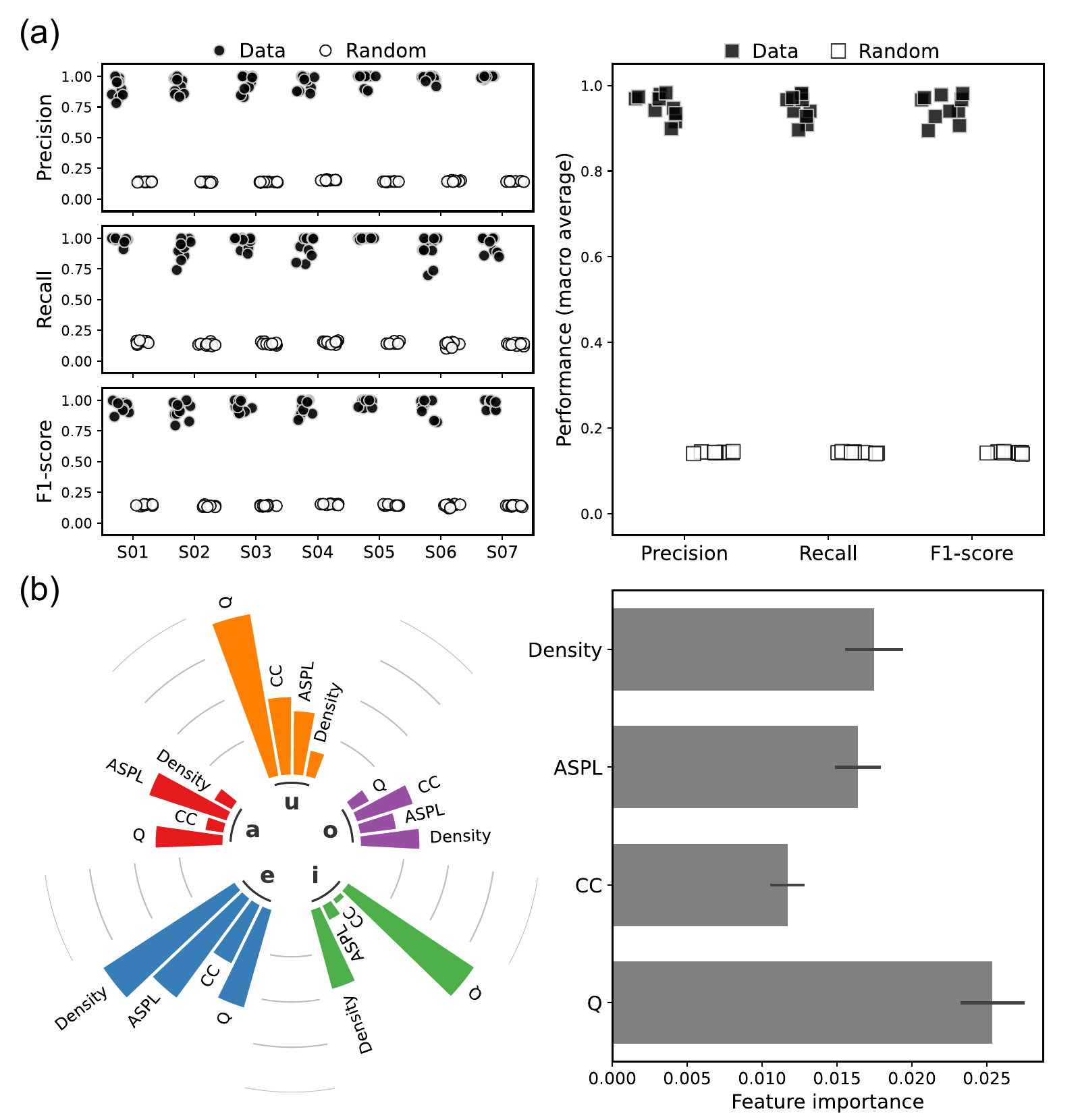}
    \caption{\label{performance} Models using visibility graphs-based metrics as features. (a) Performances of the trained models on the independent test sets, showing precision, recall, and F1-score for each class (each subject) in the left panel, and macro average performances in the right panel. (b) Feature importance analysis.}
\end{figure}

We show the feature importance analysis extracted from the Random Forest models in \Cref{performance}b, where Shapley values computed with SHAP on the test set reveal the contribution of topological metrics derived from visibility graphs to the prediction process across multiple models. The circular barplot on the left illustrates the average distribution of feature importances across models for the four graph-based metrics, i.e., link density (Density), average shortest path length (ASPL), clustering coefficient (CC), and modularity (Q), evaluated for the five vowels. Each vowel is associated with a distinct pattern, indicating that different graph metrics are more prominent for certain vowels. This vowel-dependent distribution suggests that the anatomical structure of the vocal tract influences not only the spectral characteristics of vowels but also the topology of the corresponding visibility graphs. By encoding these topological features, the models capture speaker-specific variations in vocal tract configurations, improving speaker differentiation. The bar plot on the right quantifies the average feature importance, highlighting the relative contribution of each topological metric to model performance. Among them, modularity (Q) consistently emerges as the most influential, suggesting its strong discriminative power in speaker identification tasks.

\subsection{Robustness analysis of LPC order and correlation threshold}

\begin{figure}[!ht]
    \centering
    \includegraphics[scale=0.55]{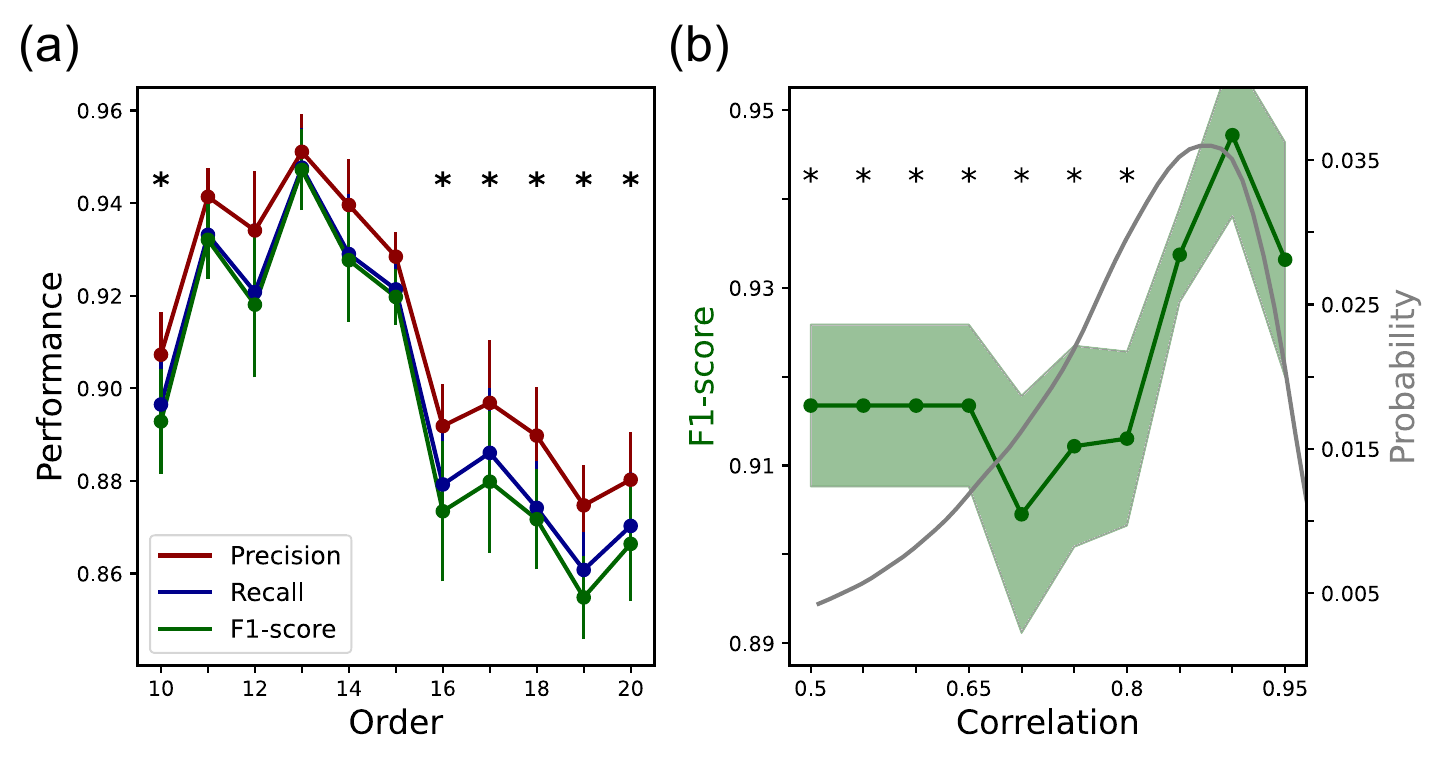}
    \caption{\label{robustness} Evaluation of models robustness. Here we show the performance of models after replying the full procedure but using different LPC orders in the computation of spectral functions (a), and using different correlation thresholds in the process of representative spectral functions selection (b). We also show the correlation probability for a better understanding of how performance is affected due to the spectral function degradation, by extracting representative samples for different correlation thresholds.}
\end{figure}

The performance of the model was evaluated across varying LPC orders and correlation thresholds to assess the robustness of our approach. \Cref{robustness}a illustrates the effect of changing the LPC order on classification performance, measured in terms of precision, recall, and F1-score. The results show that performance has consistently high values for a range of LPC orders between 11 and 15, with all three metrics (precision, recall, and F1-score) reaching their highest values at orders of 11 and 13. Asterisks indicate significant differences between the order 13 performances and lower values observed at orders below 11 and above 15 (Wilcoxon rank-sum test, $p < 0.05$). The performance drop at these orders suggests that the complexity of the spectral envelope is optimal to capture meaningful information for the order used in the main analysis. However, the plot shows that performances for other orders also remain relatively high.

In \Cref{robustness}b we present the results of varying the correlation threshold used in the binarization process during the selection of representative spectra. The F1-score remains relatively stable for correlation thresholds between 0.5 and 0.8, with values significantly below the threshold of 0.9 used in the main analysis (Wilcoxon rank-sum test, $p < 0.05$). A sharp increase in performance is observed for thresholds above 0.85, with the best F1-score achieved at a threshold of 0.9. Additionally, the gray curve in the same panel represents the probability distribution of correlations across different threshold values. This analysis indicates that our method is robust across a wide range of correlation thresholds, with relatively high performances. The optimal performance occurs for high correlation thresholds, ensuring the use of similar spectral functions as representative spectral profiles. For the threshold of 0.95 the restriction of similarity is hard enough to obtain small giant components, reflecting in a detriment of the performance.

\section{Discussion}

The results of our study underscore the potential of visibility graphs as a promising approach for speaker recognition tasks based on spectral data analysis, leveraging the topological properties embedded within visibility graphs constructed from power spectral functions. The speaker recognition models consistently achieved robust, significantly better-than-chance results on independent test data, demonstrating notably high performance. This suggests that the applied methodology—emphasizing the use of visibility graphs in the spectral domain to extract features for decision tree-based ensemble classifiers—can achieve substantial accuracy. This highlights the powerful potential of visibility graph-based feature extraction as an effective method for capturing spectral information related to vocal production differences between speakers, a finding that, to our knowledge, has not been previously reported.

Visibility graphs have already proven to be useful for capturing structural properties in various types of data, including time series \cite{lacasa_time_2008, luque_horizontal_2009} and images \cite{lacasa_visibility_2017}, particularly for peak detection \cite{nunez_detecting_2012}, while also accounting for complex relationships between amplitude patterns \cite{lacasa_visibility_2009, wen_visibility_2022}. For this reason, they are especially promising for analyzing real audio data in the frequency domain \cite{yela_spectral_2019}, where harmonic content is modulated by an envelope. Specifically, when studying human vocalizations, visibility graphs seem capable of extracting distinctive spectral features resulting from the modulation of the vocal tract acting as a passive filter \cite{titze_principles_1994}. Since vocal tract resonances depend on anatomical structures and their properties serve as unique voiceprints for individuals \cite{trevisan_topological_2005}, visibility graphs present themselves as a valuable tool for speaker recognition. In this work, we provide evidence supporting this application.

We focused our analysis on four carefully selected graph-based metrics as features, and this choice is justified as it likely contributed to the success of our approach. These metrics allow us to characterize the graph at different scales. First, link density serves as a local property, focusing on the connections of each individual node. The clustering coefficient also operates at a local scale but considers a node and the connections of its immediate neighbors. Modularity characterizes the network at a meso-scale by examining the connectivity properties within communities. Lastly, the average shortest path length offers insights at a global scale, revealing topological properties related to the overall network structure and its extension. Other commonly used metrics were considered for this analysis. For instance, the graph diameter was initially tested but discarded due to its collinearity with the average shortest path length and its limited utility due to its discretization into integer values. Similarly, the average degree—frequently used in the literature for visibility graphs—was excluded since it can be shown to be trivially identical to link density, differing only by a normalization factor, as can be demonstrated by Euler’s theorem. Although multicollinearity does not alter the performance of a Random Forest model, it can affect the inference about the feature importance, and hence last graph metrics were excluded from the analysis since they were highly correlated to other metrics in the development set, as expected. We view our selection of metrics as a meaningful first step in addressing this problem. However, future research could refine this approach by exploring more fine-tuned indices at each scale to further enhance the analysis.

To assess feature importance in prediction, we computed Shapley values on the test set using SHAP \cite{lundberg_unified_2017}, a game-theoretic approach that provides a more reliable estimation than impurity-based importance measures. Our results highlight modularity as the most significant feature (see right panel of \Cref{performance}b). The predictive power of modularity can be attributed to its ability to capture differences in spectral structures. Since spectral peaks detected by the visibility graph serve as community separators, the high importance of modularity emphasizes the role of harmonics in capturing inter-subject variability—an effect also observed in spectral function profiles (see \Cref{spectrum}).

While this study primarily focuses on demonstrating the effectiveness of visibility graph metrics for speaker recognition, feature importance analysis provides additional insights into the interpretability of this approach. However, correlation bias remains a challenge in this context. In tree-based SHAP implementations \cite{lundberg_local_2020}, the interventional method has been proposed as a way to reduce the impact of multicollinearity in Shapley value estimation, since it artificially breaks the correlation between attributes and prevents the model from compensating for the absence of a variable with other correlated variables \cite{janzing_feature_2020}. Although a more detailed assessment of correlation bias introduced during model training could further refine feature importance analysis, this lies beyond the primary scope of our work, which aims to highlight the potential of visibility graphs for speaker recognition. Nonetheless, a dedicated study focusing on the deeper analysis of feature importance and its relationship with vocal production could be a valuable direction for future research.

We achieved robust speaker identification performance using graph-based metrics. The visibility graph-derived features appear to capture higher-order structural information inherent in spectral representations, thereby enhancing the discriminative power of speaker recognition models. Additionally, the feature importance analysis allowed us to identify key spectral characteristics critical for distinguishing between speakers, facilitating the development of more interpretable and effective speaker recognition models, while deepening our understanding of the underlying phenomenology. As a result, graph-based measures could serve as a valuable complement to traditional acoustic features in audio analysis, as they capture complex relationships within spectral data, enabling a more comprehensive examination of frequency-domain information \cite{melo_graph-based_2020}. This capability not only advances speaker recognition techniques but also points to broader applications of visibility graphs in areas such as speech processing, music analysis, and audio classification.

Furthermore, our study highlights the potential of visibility graphs in tackling the challenges of speaker recognition in real-world scenarios. Random Forest, as a bagging strategy, allows the use of strong learners for complex tasks without overfitting by reducing variance through its parallel configuration \cite{breiman_random_2001}. However, the hyperparameter optimization of our models resulted in shallow estimators, indicating that even more complex classification tasks could be effectively addressed. This suggests that the scalability of our approach, particularly for distinguishing a larger number of speakers, warrants further exploration and testing. Additionally, the robustness of our method, even in the presence of representative degradation of power spectral functions, indicates its ability to perform well in diverse environmental conditions and under higher intra-subject variability. This makes it a promising candidate for practical applications such as biometric authentication systems and security surveillance.

\section{Conclusions}

This study bridges the gap between the theoretical framework of visibility graphs and the practical domain of speaker recognition. We present a novel approach wherein we apply visibility graph analysis to spectral series data obtained from recordings of male speakers articulating Spanish vowels. By transforming spectral representations into visibility graphs, we capture the underlying topological properties inherent in the frequency spectra of vocalizations. Through a comprehensive analysis leveraging graph theory metrics, we extract discriminative features crucial for speaker identification. Our approach offers a promising avenue for advancing speaker recognition technology by harnessing the power of visibility graphs to uncover intricate patterns within spectral data, thereby enhancing the accuracy and robustness of biometric systems in real-world applications.

In the realm of biometrics, speaker recognition stands as a significant challenge with far-reaching implications in cybersecurity and identity verification. In this framework, the importance of developing robust speaker recognition systems capable of distinguishing between individuals accurately and efficiently becomes a clear short-term need. With recent advancements in artificial intelligence, the exploration for reliable biometric systems has become overriding, especially in light of emerging threats such as AI-generated voice impersonations. Therefore, there exists a pressing necessity of innovative methodologies that can enhance the speaker identification. In that sense, our approach boards the problem with a non-previously used strategy. Further research is needed to define scope and limitations of this approach, but also to consider if topological properties of graph-based representation on the spectral domain, as used in this work, could provide a complementary point of view for the boosting of current techniques.

\subsection*{CRediT authorship contribution statement}

\hspace{6pt} \textbf{Hernan Bocaccio:} Data Curation, Conceptualization, Methodology, Software, Writing - Original Draft, Writing - Review \& Editing, Visualization. \textbf{Sergio Iglesias-Pérez:} Investigation, Writing - Review \& Editing. \textbf{Miguel Romance:} Investigation, Writing - Review \& Editing. \textbf{Regino Criado:} Investigation, Writing - Review \& Editing. \textbf{Gabriel Bernardo Mindlin:} Conceptualization, Methodology, Project Administration, Software, Supervision, Writing - Original Draft, Writing - Review \& Editing, Visualization.

\subsection*{Declaration of Competing Interest}

\hspace{6pt} The authors declare no conflict of interests.

\subsection*{Acknowledgments}

\hspace{6pt} This work has been partially supported by INCIBE/URJC Agreement M3386/2024/0031/001 within the framework of the Recovery, Transformation and Resilience Plan funds of the European Union (Next Generation EU); and by Consejo Nacional de Investigaciones Científicas y Técnicas (CONICET). HB is a postdoctoral fellow from CONICET. GBM is grateful for the hospitality of URJC during his sabbatical stays.

\subsection*{Data accessibility}

\hspace{6pt} Data and codes required to reproduce this article are available at a public repository (https://github.com/HBocaccio/On-the-application-of-Visibility-Graphs-in-the-Spectral-Domain-for-Speaker-Recognition).


\begin{thebibliography}{10}

\bibitem{lacasa_time_2008}
L.~Lacasa, B.~Luque, F.~Ballesteros, J.~Luque, and J.~C. Nuño, ``From time series to complex networks: {The} visibility graph,'' {\em Proceedings of the National Academy of Sciences}, vol.~105, no.~13, pp.~4972--4975, 2008.

\bibitem{lacasa_visibility_2009}
L.~Lacasa, B.~Luque, J.~Luque, and J.~C. Nuño, ``The visibility graph: {A} new method for estimating the {Hurst} exponent of fractional {Brownian} motion,'' {\em Europhysics Letters}, vol.~86, no.~3, p.~30001, 2009.

\bibitem{luque_horizontal_2009}
B.~Luque, L.~Lacasa, F.~Ballesteros, and J.~Luque, ``Horizontal visibility graphs: {Exact} results for random time series,'' {\em Physical Review E}, vol.~80, no.~4, p.~046103, 2009.

\bibitem{partida_chaotic_2022}
A.~Partida, S.~Gerassis, R.~Criado, M.~Romance, E.~Giráldez, and J.~Taboada, ``The chaotic, self-similar and hierarchical patterns in {Bitcoin} and {Ethereum} price series,'' {\em Chaos, Solitons \& Fractals}, vol.~165, p.~112806, 2022.

\bibitem{iglesias-perez_increasing_2022}
S.~Iglesias-Perez and R.~Criado, ``Increasing the {Effectiveness} of {Network} {Intrusion} {Detection} {Systems} ({NIDSs}) by {Using} {Multiplex} {Networks} and {Visibility} {Graphs},'' {\em Mathematics}, vol.~11, no.~1, p.~107, 2022.

\bibitem{iglesias-perez_temporal_2023}
S.~Iglesias-Perez and R.~Criado, ``Temporal metagraph: {A} new mathematical approach to capture temporal dependencies and interactions between different entities over time,'' {\em Chaos, Solitons \& Fractals}, vol.~175, p.~113940, 2023.

\bibitem{iglesias-perez_combining_2023}
S.~Iglesias-Perez, S.~Moral-Rubio, and R.~Criado, ``Combining multiplex networks and time series: {A} new way to optimize real estate forecasting in {New} {York} using cab rides,'' {\em Physica A: Statistical Mechanics and its Applications}, vol.~609, p.~128306, 2023.

\bibitem{zheng_visibility_2021}
M.~Zheng, S.~Domanskyi, C.~Piermarocchi, and G.~I. Mias, ``Visibility graph based temporal community detection with applications in biological time series,'' {\em Scientific Reports}, vol.~11, no.~1, p.~5623, 2021.

\bibitem{ahmadlou_visibility_2012}
M.~Ahmadlou and H.~Adeli, ``Visibility graph similarity: {A} new measure of generalized synchronization in coupled dynamic systems,'' {\em Physica D: Nonlinear Phenomena}, vol.~241, no.~4, pp.~326--332, 2012.

\bibitem{melo_graph-based_2020}
D.~d. F.~P. Melo, I.~d.~S. Fadigas, and H.~B. d.~B. Pereira, ``Graph-based feature extraction: {A} new proposal to study the classification of music signals outside the time-frequency domain,'' {\em PLOS ONE}, vol.~15, no.~11, p.~e0240915, 2020.

\bibitem{sulaimany_visibility_2023}
S.~Sulaimany and Z.~Safahi, ``Visibility graph analysis for brain: scoping review,'' {\em Frontiers in Neuroscience}, vol.~17, 2023.

\bibitem{lacasa_visibility_2017}
L.~Lacasa and J.~Iacovacci, ``Visibility graphs of random scalar fields and spatial data,'' {\em Physical Review E}, vol.~96, no.~1, p.~012318, 2017.

\bibitem{iacovacci_visibility_2020}
J.~Iacovacci and L.~Lacasa, ``Visibility {Graphs} for {Image} {Processing},'' {\em IEEE Transactions on Pattern Analysis and Machine Intelligence}, vol.~42, no.~4, pp.~974--987, 2020.

\bibitem{pei_texture_2021}
L.~Pei, Z.~Li, and J.~Liu, ``Texture classification based on image (natural and horizontal) visibility graph constructing methods,'' {\em Chaos (Woodbury, N.Y.)}, vol.~31, no.~1, p.~013128, 2021.

\bibitem{wen_visibility_2022}
T.~Wen, H.~Chen, and K.~H. Cheong, ``Visibility graph for time series prediction and image classification: a review,'' {\em Nonlinear Dynamics}, vol.~110, no.~4, pp.~2979--2999, 2022.

\bibitem{yela_spectral_2019}
D.~F. Yela, D.~Stowell, and M.~Sandler, ``Spectral {Visibility} {Graphs}: {Application} to {Similarity} of {Harmonic} {Signals},'' in {\em 2019 27th {European} {Signal} {Processing} {Conference} ({EUSIPCO})}, pp.~1--5, 2019.

\bibitem{nunez_detecting_2012}
A.~Nuñez, L.~Lacasa, E.~Valero, J.~P. Gómez, and B.~Luque, ``Detecting series periodicity with horizontal visibility graphs,'' {\em International Journal of Bifurcation and Chaos}, vol.~22, no.~07, p.~1250160, 2012.

\bibitem{fant_acoustic_1971}
G.~Fant, {\em Acoustic {Theory} of {Speech} {Production}: {With} {Calculations} {Based} on {X}-{Ray} {Studies} of {Russian} {Articulations}}.
\newblock Walter de Gruyter, 1971.

\bibitem{lieberman_speech_1988}
P.~Lieberman and S.~E. Blumstein, {\em Speech {Physiology}, {Speech} {Perception}, and {Acoustic} {Phonetics}}.
\newblock Cambridge University Press, 1988.

\bibitem{titze_principles_1994}
I.~R. Titze, {\em Principles of {Voice} {Production}}.
\newblock Prentice Hall, 1994.

\bibitem{trevisan_topological_2005}
M.~A. Trevisan, M.~C. Eguia, and G.~B. Mindlin, ``Topological voiceprints for speaker identification,'' {\em Physica D: Nonlinear Phenomena}, vol.~200, no.~1, pp.~75--80, 2005.

\bibitem{sainburg_finding_2020}
T.~Sainburg, M.~Thielk, and T.~Q. Gentner, ``Finding, visualizing, and quantifying latent structure across diverse animal vocal repertoires,'' {\em PLOS Computational Biology}, vol.~16, no.~10, p.~e1008228, 2020.

\bibitem{atal_speech_1971}
B.~S. Atal and S.~L. Hanauer, ``Speech {Analysis} and {Synthesis} by {Linear} {Prediction} of the {Speech} {Wave},'' {\em The Journal of the Acoustical Society of America}, vol.~50, no.~2B, pp.~637--655, 1971.

\bibitem{mcfee_librosa_2015}
B.~McFee, C.~Raffel, D.~Liang, D.~P. Ellis, M.~McVicar, E.~Battenberg, and O.~Nieto, ``librosa: {Audio} and music signal analysis in python.,'' in {\em {SciPy}}, pp.~18--24, 2015.

\bibitem{virtanen_scipy_2020}
P.~Virtanen, R.~Gommers, T.~E. Oliphant, M.~Haberland, T.~Reddy, D.~Cournapeau, E.~Burovski, P.~Peterson, W.~Weckesser, and J.~Bright, ``{SciPy} 1.0: fundamental algorithms for scientific computing in {Python},'' {\em Nature methods}, vol.~17, no.~3, pp.~261--272, 2020.

\bibitem{rubinov_complex_2010}
M.~Rubinov and O.~Sporns, ``Complex network measures of brain connectivity: {Uses} and interpretations,'' {\em NeuroImage}, vol.~52, no.~3, pp.~1059--1069, 2010.

\bibitem{lan_fast_2015}
X.~Lan, H.~Mo, S.~Chen, Q.~Liu, and Y.~Deng, ``Fast transformation from time series to visibility graphs,'' {\em Chaos: An Interdisciplinary Journal of Nonlinear Science}, vol.~25, no.~8, p.~083105, 2015.

\bibitem{yela_online_2020}
D.~F. Yela, F.~Thalmann, V.~Nicosia, D.~Stowell, and M.~Sandler, ``Online visibility graphs: {Encoding} visibility in a binary search tree,'' {\em Physical Review Research}, vol.~2, no.~2, p.~023069, 2020.

\bibitem{breiman_random_2001}
L.~Breiman, ``Random {Forests},'' {\em Machine Learning}, vol.~45, no.~1, pp.~5--32, 2001.

\bibitem{lundberg_unified_2017}
S.~M. Lundberg and S.-I. Lee, ``A unified approach to interpreting model predictions,'' in {\em Proceedings of the 31st {International} {Conference} on {Neural} {Information} {Processing} {Systems}}, {NIPS}'17, pp.~4768--4777, Curran Associates Inc., 2017.

\bibitem{lundberg_local_2020}
S.~M. Lundberg, G.~Erion, H.~Chen, A.~DeGrave, J.~M. Prutkin, B.~Nair, R.~Katz, J.~Himmelfarb, N.~Bansal, and S.-I. Lee, ``From local explanations to global understanding with explainable {AI} for trees,'' {\em Nature Machine Intelligence}, vol.~2, no.~1, pp.~56--67, 2020.

\bibitem{janzing_feature_2020}
D.~Janzing, L.~Minorics, and P.~Bloebaum, ``Feature relevance quantification in explainable {AI}: {A} causal problem,'' in {\em Proceedings of the {Twenty} {Third} {International} {Conference} on {Artificial} {Intelligence} and {Statistics}}, pp.~2907--2916, PMLR, 2020.
\newblock ISSN: 2640-3498.

\end{thebibliography}

\end{document}